\documentclass[12pt]{iopart}

\usepackage{graphicx}
\usepackage{amsmath}
\usepackage{xcolor}
\usepackage{soul}
\usepackage[natbib=true,style=numeric,sorting=none]{biblatex}
\begin{document}

\title[SOLPS-ITER simulations of detachment in MAST-U Super-X]{Comparison between MAST-U conventional and Super-X configurations through SOLPS-ITER modelling}


\author{A. Fil$^{1,2}$, B. Lipschultz$^2$, D. Moulton$^1$, A. Thornton$^1$, \\B. D. Dudson$^{2,3}$, O. Myatra$^{1,2}$, K. Verhaegh$^1$, and the EUROFusion MST1 team$^*$}

\address{$^1$ United Kingdom Atomic Energy Authority, Culham Centre for Fusion Energy, Culham Science Centre, Abingdon, Oxon, OX14 3DB, UK}

\address{$^2$ York Plasma Institute, University of York, Heslington, York, YO10 5DD, UK}

\address{$^3$ Lawrence Livermore National Laboratory, 7000 East Ave, Livermore, CA 94550, USA}

\address{$^*$ See author list of “B. Labit et al 2019 Nucl. Fusion 59 086020 (https://doi.org/10.1088/1741-4326/ab2211)}

\ead{alexandre.fil@ukaea.uk}

\begin{abstract}
MAST-U has recently started operating with a Super-X divertor, designed to increase total flux expansion and neutral trapping, both predicted through simple analytic models and SOLPS calculations to reduce the plasma and impurity density detachment thresholds. In this study, utilising the SOLPS-ITER code, we are quantifying the possible gain allowed by the MAST-U Super-X and neutral baffling geometry, in terms of access to detachment. We show that a significant reduction of the upstream density detachment threshold (up to a factor 1.6) could be achieved in MAST-U, for the Super-X, as opposed to conventional divertor geometry, mainly through an increased total flux expansion, neutral trapping being found very similar between the different configurations. We also show that variations of the strike-point angle are complex to interpret in such a tightly baffled geometry, and that a case in which the target normal points more towards the separatrix does not necessarily imply a lower detachment threshold. As in previous calculations for TCV, we quantify the role of neutral effects through developing and applying a quantitative definition of neutral trapping.
\end{abstract}

\maketitle

\section{Introduction}
For future fusion reactors, the process of divertor detachment will be required to reduce the power to the divertor targets and keep it low enough to prevent excessive material erosion and/or degradation, that would undermine efforts to obtain economically viable devices. In the conventional magnetic configuration that is currently foreseen for DEMO \cite{maviglia_effect_2017}, one would need to raise the total plasma radiation to values of over $95\%$ to prevent melting in the divertor \cite{kallenbach_impurity_2013}\cite{wischmeier_high_2015}. Many alternative divertor magnetic topologies are currently being studied to tackle this issue, through experiments \cite{theiler_results_2017}\cite{petrie_effect_2013}\cite{soukhanovskii_snowflake_2011}\cite{reimerdes_tcv_2017}, simple analytic models \cite{petrie_effect_2013}\cite{lipschultz_sensitivity_2016}\cite{kotschenreuther_super_2010} and numerical models \cite{moulton_using_2017}\cite{umansky_analysis_2009}\cite{fil_separating_2020}. Compared to the conventional divertor designs \cite{groth_divertor_2015}\cite{jaervinen_impact_2016}\cite{loarte_effects_2001}\cite{loarte_plasma_1998}, those alternative concepts primarily include variations of the magnetic topology, targeting an easier access to detachment and enhanced power losses in the divertor. Maximizing neutral trapping through divertor closure is another path for optimizing the divertor \cite{casali_neutral_2020}\cite{fevrier_divertor_2021}\cite{reimerdes_initial_2021}. \\
MAST-U \cite{morris_mast_2014} has recently started operating with such an alternative divertor, the Super-X divertor. Several effects could be at play to lead to a reduction in detachment threshold in the Super-X compared to more conventional divertor topologies, and the modelling presented in this paper aims to highlight some of them. The first effect, which has been thoroughly studied, is the increase of total flux expansion when the strike point is moved to a region with lower magnetic field, through the increase of the cross-sectional area of the flux tubes \cite{petrie_effect_2013}\cite{lipschultz_sensitivity_2016}\cite{moulton_using_2017}\cite{stangeby_plasma_2000}. Theoretically, this should reduce the parallel heat flux as $q_{t,\parallel} \propto |B_{tot, target}| \propto \frac{1}{R_t}$ (as $|B_{tot}| \simeq |B_T|$ in most cases). This effect has also been observed in recent modelling, but can only be isolated when another effect on the divertor plasma, that of neutral effects (or neutral trapping) are the same between the different configurations \cite{moulton_using_2017}\cite{fil_separating_2020}. However, it is usually quite challenging to keep such neutral effects constant experimentally (or even in modeling), as shown by recent work on TCV \cite{theiler_results_2017}\cite{fil_separating_2020}, and differences in neutral trapping between configurations can either counteract or amplify the effect of total flux expansion. \\
With MAST-U being so tightly baffled, one can expect neutral trapping to be enhanced in the Super-X chamber (in contrast to TCV cases with high total flux expansion) and thus an even greater reduction of the target temperature (or of the detachment threshold) than the one predicted assuming similar neutral trapping in both configurations (as in \cite{lipschultz_sensitivity_2016}). \\
The benefits of the MAST-U Super-X configuration have already been demonstrated through modelling \cite{rozhansky_modeling_2013}\cite{havlickova_solps_2015}, and will be further studied in this paper, looking at the changes in the density detachment threshold (i.e. the upstream density at which a rollover of the total target ion flux is observed) between Conventional and Super-X configurations, their neutral trapping properties, as well as the influence of the strike point angle to the target surface for the Super-X configuration. Section \ref{sec:Methods} will present the code used (SOLPS-ITER) and the input parameters, while section \ref{sec:results} will present the results of the modelling. We will then discuss the results and conclude in section \ref{sec:conclusions}.

\section{Methods}
\label{sec:Methods}
\begin{figure}[!ht]
\centering
\includegraphics[width=13cm]{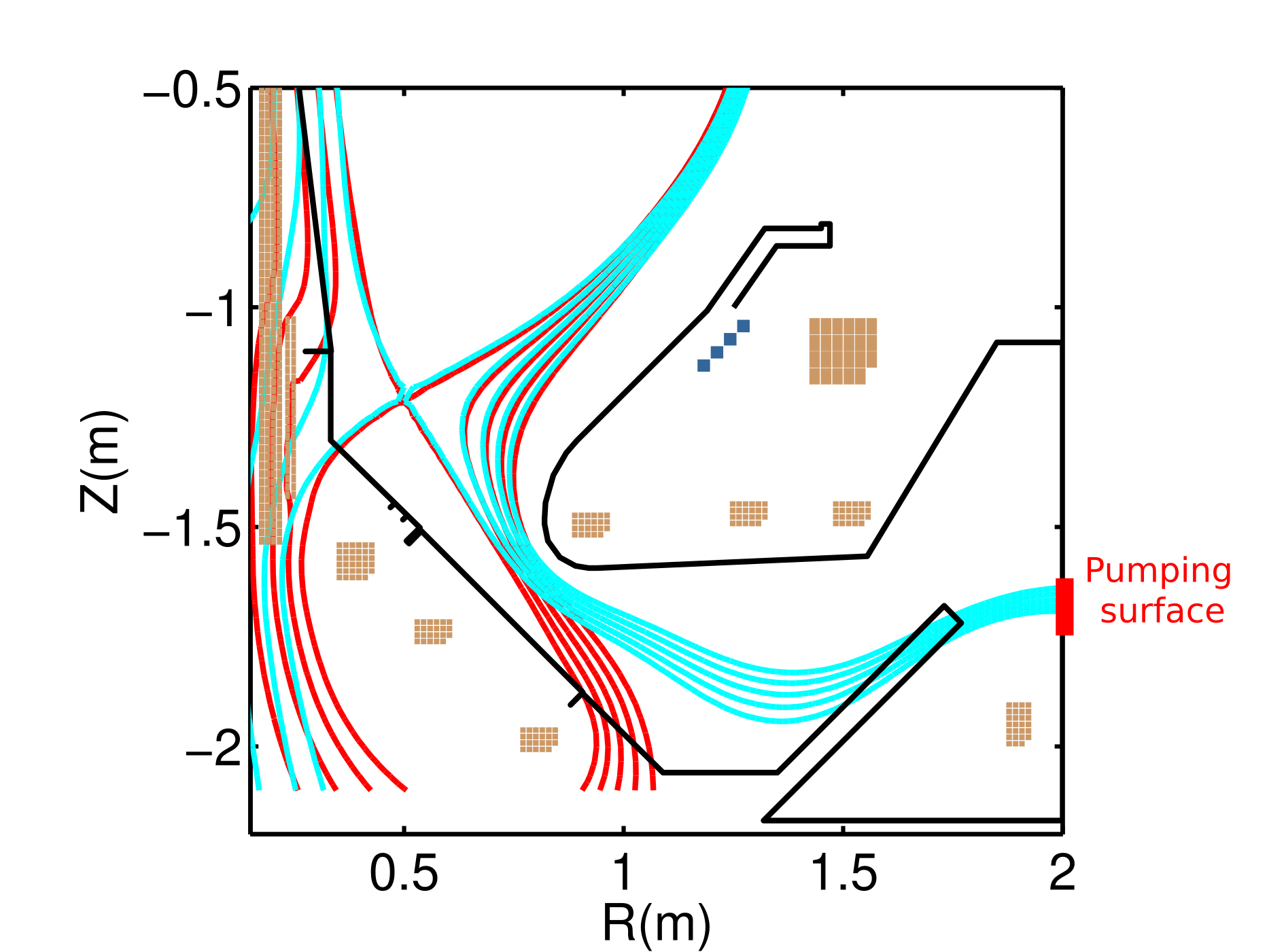}
\caption{Set of equilibria generated for MAST-U experiments to study the effect of total flux expansion on detachment (Courtesy of A. Thornton). Special care has been taken to keep the X-point position and poloidal flux expansion similar between the configurations. Also plotted is the position of the realistic pumping surface that will be used in the simulations.}
\label{fig:mastu} 
\end{figure}

In this study, we use the code SOLPS-ITER \cite{wiesen_new_2015}\cite{bonnin_presentation_2016} to model 3 different MAST-U configurations. Two of those three configurations are shown in Figure \ref{fig:mastu} (the third configuration for these studies is introduced later); the red configuration is designated "Conventional" (even though the outer strike point is already well inside the divertor chamber) and the light blue configuration is designated "Super-X". The latter is a variation of the Super-X configuration that had been studied in several papers \cite{havlickova_solps_2015}\cite{moulton_eps_2017}. It has a lower poloidal flux expansion and no extra-null in the poloidal magnetic field. Having similar poloidal flux expansion facilitates the comparison between the different configurations. Note that both equilibria are top-down symmetric double nulls.\\

\begin{figure}[!ht]
\centering
\includegraphics[width=15.5cm]{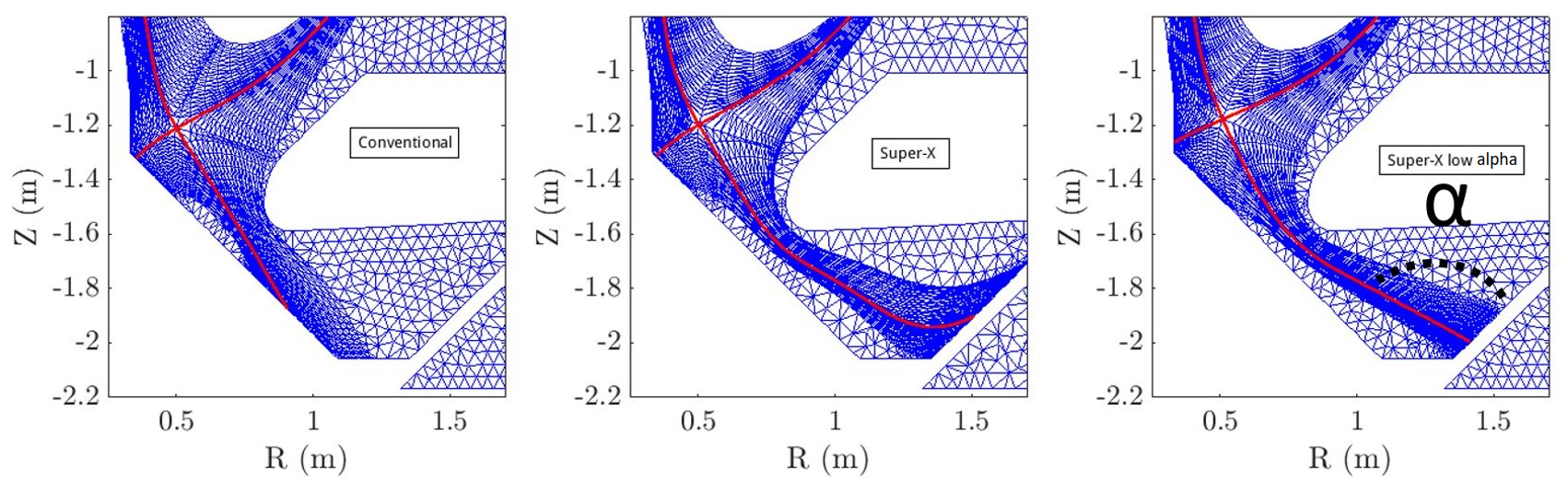}
\caption{Set of MAST-U SOLPS-ITER grids for the "Conventional" and "Super-X" configurations. The separatrices are shown in red and the grid points in blue. The plasma domain corresponds to the flux-aligned grid, while the triangular grid is only for the kinetic neutrals simulated with EIRENE \cite{reiter_progress_1992}. The Super-X low $\alpha$ case reduces the strike point angle of the Super-X from $162^{\circ}$ to $105^{\circ}$ (definition of the angle $\alpha$ is shown on the Figure).}
\label{fig:grids} 
\end{figure}

The corresponding SOLPS-ITER grids are shown in Figure \ref{fig:grids}, focused on the lower divertor region. The comparison of the Conventional vs. the Super-X, in terms of detachment access, will be presented in section \ref{sec:super}. Also shown is the third computational grid corresponding to a variation of the Super-X configuration, having a lower angle between the outer strike points and the target normal. The comparison of the two Super-X configurations will be presented in section \ref{sec:beta}.

In all the following simulations, we run in toroidal geometry and are not including drift effects. Drift effects can give rise to bifurcation and redistribution of power and
particles between the targets \cite{jaervinen_PRL}. We believe this should not affect strongly the qualitative trends observed during the presented scans but could affect the quantitative values. Recent MAST-U SOLPS-ITER simulations have found minimal impact of drifts on the detachment threshold, but for a lower input power than in the simulations presented in this paper.
We use the same fixed ad-hoc coefficients for the perpendicular particle and heat transport ($D_{\perp} = 0.2$ $m^2 s^{-1}$, $\kappa_{\perp, e,i} = 1$ $m^2 s^{-1}$) that were used in TCV, and a fixed input power ($P_{input}=2.5$ $MW$). We set the power entering the grid from the core to be $P_{input}$, and zero flux core boundary condition for the densities, i.e. the flux of ions leaving the core is equal and opposite to the inflow of deuterium neutral particles. We also model Carbon physical and chemical sputtering of the first wall, with a chemical sputtering yield of $3 \%$. The pumps (i.e pumping surfaces) are placed at the end of the Super-X chambers (see Figure \ref{fig:mastu}). The recycling coefficient $R=0.98955$ is consistent with the pumping speed of the turbo pumps available during the first physics campaign of MAST-U. In attached conditions, we obtain $\lambda_q \simeq 4.5$ mm in the simulations (as calculated in \cite{pitts_physics_2019}).\\ 
As these simulations were done before the first MAST-U campaign, and at much higher input power, they can't be compared to experiments as of yet. Work is on-going to reproduce the experimental results of the first campaign, but will be presented in a separate publication.
Finally, wall pumping is not taken into account, but can significantly change the results (i.e. large extra sink for the neutrals) and will need to be properly estimated in the experiments to guide future modelling. While the change of those parameters can affect each simulation, they do not seem to impact significantly the ratio of detachment threshold between configurations (similarly to what has been found in TCV modeling \cite{fil_separating_2020}).

\section{Results}
\label{sec:results}

\subsection{Conventional vs. Super-X}
\label{sec:super}

\begin{figure}[!ht]
\centering
\includegraphics[width=15cm]{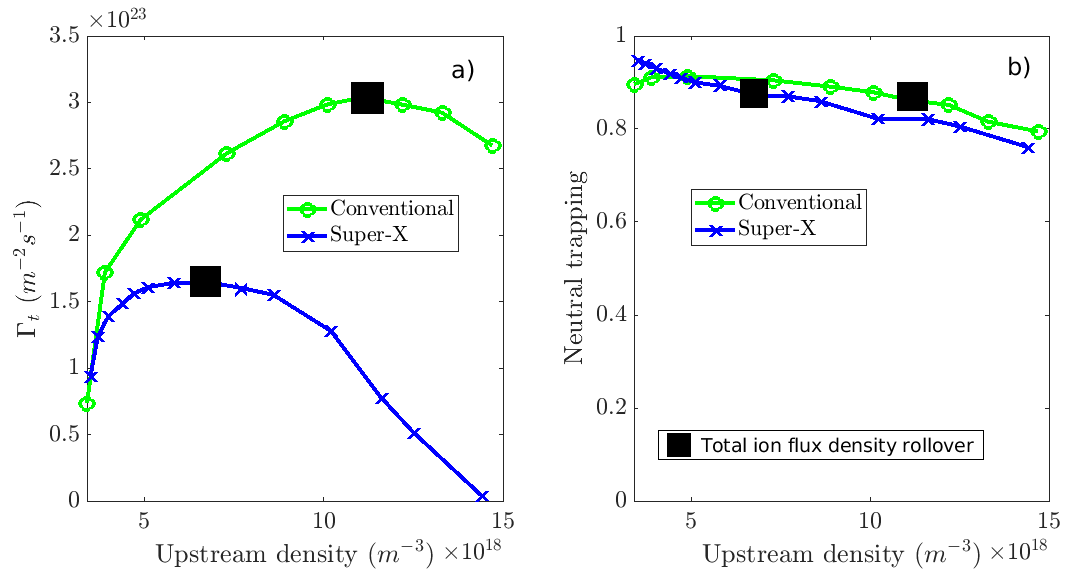}
\caption{SOLPS-ITER predictions of the MAST-U outer lower target a) total ion flux density and b) associated neutral trapping, for a Conventional and a Super-X configuration. The neutral trapping is defined in equation \ref{eq:eta}. The filled black squares are at the point of maximum outer divertor target current densities and corresponds to the upstream density detachment thresholds, with an uncertainty of $\pm 10\%$ because of the limited number of simulations.}
\label{fig:flux} 
\end{figure}

A scan of upstream density, through a scan of gas puffing rate, is performed for each configuration. The puffing location is at the inner midplane, and the gas puff rate is varied between $1\cdot10^{21}$ and $1.8\cdot10^{22}$ $D_2$ molecules per second.
\\
If total flux expansion was the only difference between the two configurations, the modified 2 point model \cite{petrie_effect_2013,lipschultz_sensitivity_2016} would predict that the ratio of detachment threshold in upstream density ($n_{u,detach}$) between the two configurations should be:
\begin{equation}
\begin{split}
        N_{thres,2PM}=\frac{n_{u, detach, CD}}{n_{u, detach, SXD}} \simeq \frac{B_{tar,CD}}{B_{tar,SXD}} \\
     \simeq \frac{R_{tar,SXD}}{R_{tar,CD}}\simeq 1.7
\end{split}
     \label{eq:nthresh}
\end{equation}
Where $B_{tar,CD}$ and $B_{tar,SXD}$ are the total magnetic fields at the CD (Conventional Divertor) and SXD (Super-X Divertor) strike points respectively. As noted, the ratio of the total magnetic field amplitudes at the strike points can be approximated as the inverse ratio of the two outer strike point radii. Note that equation \ref{eq:nthresh} assumes that the target temperature required to detach is independent of flux expansion.
\\
Figure \ref{fig:flux}a) shows the target ion flux density averaged over the whole outer lower target (including the private flux region, or PFR) for each converged simulation, versus the corresponding upstream density (defined as the density in the first SOL flux tube or grid cell, at the outer midplane). The detachment threshold, indicated by the black squares, is defined as the upstream density at the rollover point (maximum current) of the total target ion flux density (defined as $\frac{\sum_i \Gamma_{end,i}}{A}$, where $\Gamma_{end,i}$ is the ion flux in particle per second at the end of the i$^{th}$ flux tube, i.e. at the outer target, and A the outer target area). We find that the SOLPS-ITER modelling leads to an $N_{thres,SOLPS} \simeq 1.6$, similar to the analytic model predictions ($ N_{thres,2PM} \simeq 1.7$). Due to the limited number of simulations, we estimate an uncertainty on the ion flux rollover "measurement" of about $10 \%$. In other words, using the Super-X configuration allows to reduce the density detachment threshold by a factor of $\simeq 1.6$ $[\pm 15\%]$ compared to the conventional configuration for the MAST-U modelled outer lower target; this result is consistent with previous modelling of MAST-U \cite{moulton_eps_2017}. One can also calculate the ratio of detachment thresholds for the two configurations using other characteristics of the onset of detachment, such as the upstream density at which the CIII emissivity front or the ionization source detach from the target \cite{theiler_results_2017}\cite{fil_separating_2020}; such analysis provide similar $N_{thres}$ between the two configurations as found using the rollover point in the ion target flux density. 

\begin{figure}[!ht]
\centering
\includegraphics[width=15.5cm]{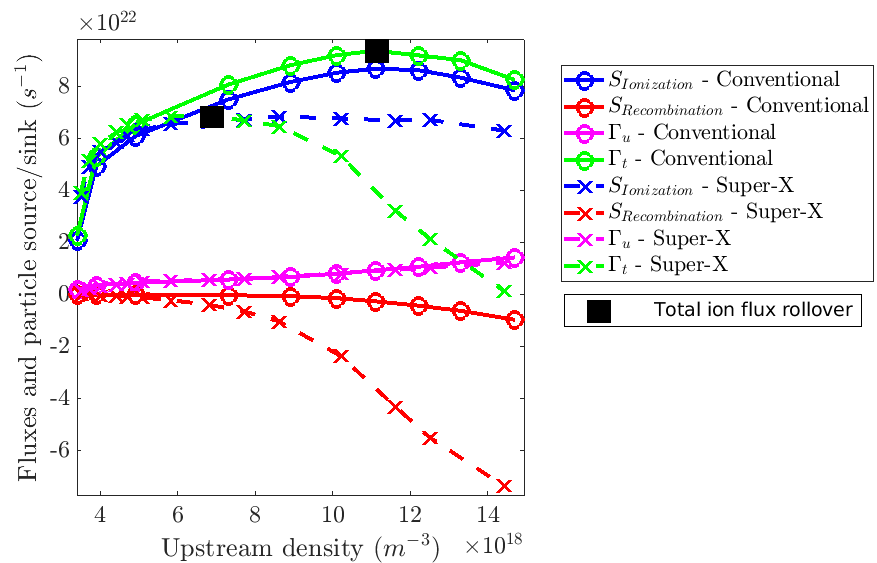}
\caption{MAST-U particle balance in the lower outer divertor for the Conventional and the Super-X configurations. $S_{ion}$ and $S_{rec}$ are, respectively, the divertor-integrated ionization source and recombination sink. $\Gamma_u$ is the upstream ion flux entering the divertor. $\Gamma_t$ is the total target ion flux. Radial transport out of the domain is negligible and not plotted.}
\label{fig:partbal} 
\end{figure}

We have also analyzed the various particle sources and sinks that contribute to the particle balance, and thus determine the ion target flux density - see Equation \ref{eq:bal} below.
\begin{equation}
     \Gamma_{target} = \Gamma_{u} + S_{ion} + S_{rec} + \Gamma_{Rad. transp.}
     \label{eq:bal}
\end{equation}
Figure \ref{fig:partbal} displays those various sources and sinks for the cases of Figure \ref{fig:flux}a), for the entire lower outer divertor domain (including the PFR). $S_{ion}$ and $S_{rec}$ are the integrated (over the lower outer divertor grid below the X-point) ionization source and recombination sink. The flux of particles into the divertor, $\Gamma_u$, and to the target, $\Gamma_t$, are also shown. In both cases the ionization source saturates as the upstream density is increased. The recombination sink plays a significant role for both cases with strong increases corresponding to when the temperature decreases below $\approx 1$ eV at the target. Ionization and recombination are thus the two dominant mechanisms responsible for the rollover of the target ion flux as opposed to $\Gamma_u$. In all the simulations, the net radial transport out of the plasma domain is orders of magnitude lower than those two sources/sinks, and thus not plotted here.

\begin{figure}[!ht]
\centering
\includegraphics[width=15cm]{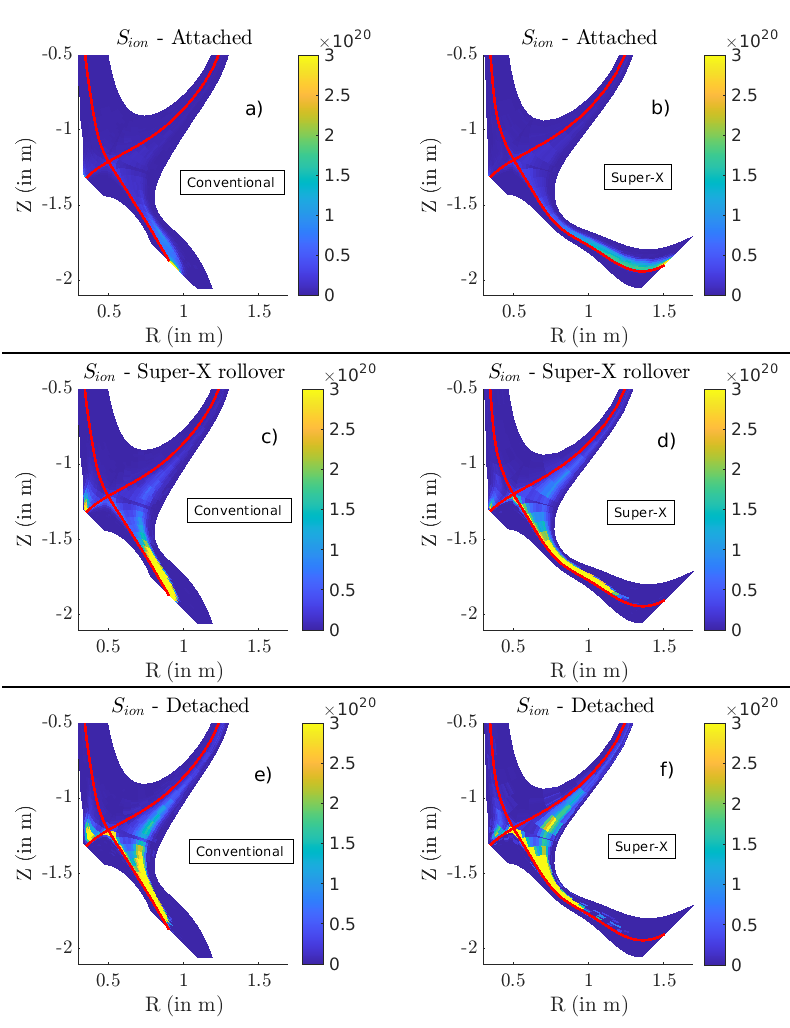}
\caption{SOLPS-ITER 2D plots of the ionization source $S_{ion}$ evolution for both configurations, with the ionization front detaching from the target as detachment progresses. In a) and b) both are in attached conditions (i.e. the upstream density is $n_{e,up} \sim 3.5\times 10^{18} m^{-3}$). c) and d) are the conventional and Super-X at the Super-X rollover point ($n_{e,up} \sim 7\times 10^{18} m^{-3}$) indicated on Figure \ref{fig:flux}. In e) and f) both are in detached conditions ($n_{e,up} \sim 1.45\times 10^{19} m^{-3}$). }
\label{fig:ion2D} 
\end{figure}

\begin{figure}[!ht]
\centering
\includegraphics[width=15cm]{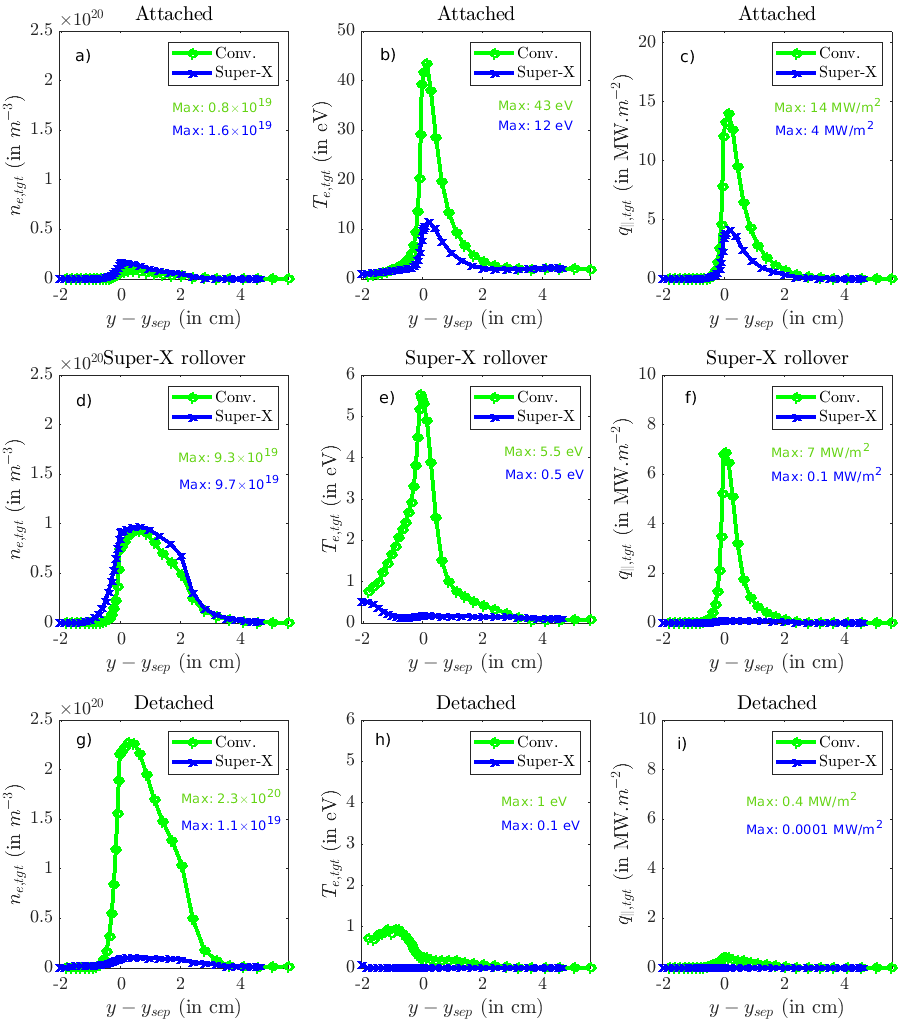}
\caption{Electron density, electron temperature and parallel heat flux profiles at the lower outer target for both configurations: a), b) and c) when both are attached ($n_{e,up} \sim 3.5\times 10^{18} m^{-3}$). d), e) and f) at the Super-X rollover point ($n_{e,up} \sim 7\times 10^{18} m^{-3}$) indicated on Figure \ref{fig:flux}. g), h) and i) when both are in detached conditions ($n_{e,up} \sim 1.45\times 10^{19} m^{-3}$).}
\label{fig:tgt} 
\end{figure}

Figure \ref{fig:ion2D} shows the 2D evolution of the ionization source at different levels of detachment for both configurations, confirming that the ionization front detaches from the target at lower upstream density for the Super-X. This figure also shows that, as detachment progresses, more ionization occurs out of the divertor domain. However, compared to TCV, this effect is limited and most of the ionization source stays inside the divertor. This illustrates the very good neutral trapping properties of MAST-U in both configurations. Figure \ref{fig:tgt} shows the lower outer target profiles of electron density, electron temperature, and parallel heat flux for both configurations at the same three different upstream densities than in Figure \ref{fig:ion2D}. As the upstream density is increased and detachment progresses, the target density increases and both the target temperature and the parallel heat flux decrease dramatically. At the same upstream density, the Super-X has significantly lower target temperature and parallel heat flux than the Conventional. Interestingly, we also observe a rollover of the target electron density when the Super-X is deeply detached (see Fig.6.d and 6.g).

\begin{figure}[!ht]
\centering
\includegraphics[width=12cm]{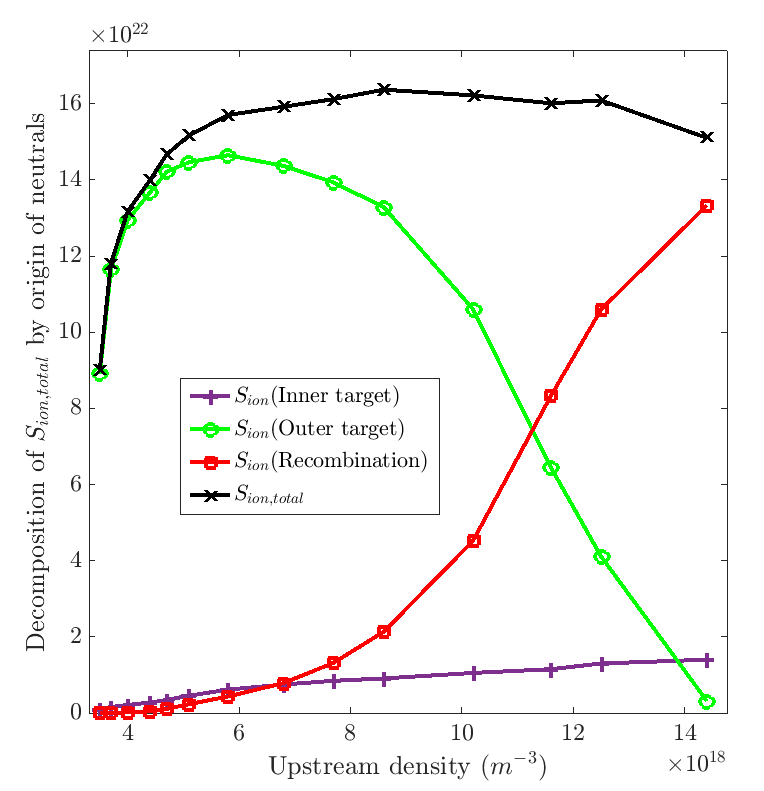}
\caption{Decomposition of the ionization source in the lower outer divertor, $S_{ion,total}$, by the origin of the neutrals getting ionized. Most of the ionization in the lower outer divertor comes from neutrals originating at the lower outer target, from the lower inner target, and from the recombination process.}
\label{fig:Siondecomp} 
\end{figure}

Going back to the particle balance, we further decompose $S_{ion}$ in Figure \ref{fig:Siondecomp}, for the Super-X configuration, in order to understand the saturation of the ionization source. This decomposition displays the origin of the neutrals which ionize in the lower outer divertor. $S_{ion,total}$ is initially dominated by the ionization of neutrals originating from the lower outer target "$S_{ion}$(Outer target)". However, "$S_{ion}$(Outer target)" rolls-over and is progressively replaced by the ionization of neutrals created by recombination processes "$S_{ion}$(Recombination)". Some neutrals originating at the lower inner target "$S_{ion}$(Inner target)" also contribute to the lower outer divertor ionization source, but account for less than $10\%$ of the total. Interestingly, $S_{ion}$(Recombination) $\simeq -S_{rec}$, i.e. a significant fraction of the recombination neutral source is re-ionised in the lower outer divertor. Only a few percents of those neutrals are getting pumped, and a few percents of them are getting ionized above the X-point and in the inner target. As detachment progresses, more and more of these recombined neutrals manage to leave the divertor domain and get ionized above the X-point (also seen on Figure \ref{fig:ion2D}), explaining partly the increase of the upstream ion flux, $\Gamma_u$, observed on Figure \ref{fig:partbal}. In other words, we have a virtual (recombining) target, where the recombined neutrals do get re-ionised (upstream of the recombination region) but, overall, don't contribute to the target flux. There is minimal power starvation and mostly power limitation (i.e. the total ionization saturates but does not drop) determining the target current.

The fact that the reduction of the detachment threshold between Conventional and Super-X is close to the model prediction (equation \ref{eq:nthresh}) suggests that neutral effects are similar between the two configurations, as was shown in \cite{fil_separating_2020}. This is further investigated in figure \ref{fig:flux}b), which shows the evolution of the neutral trapping parameter, $\eta_{RI,rec}$, throughout the upstream density scan. This parameter, first introduced in \cite{fil_separating_2020}, is used to characterize the neutral trapping properties of the different configurations. Equation \ref{eq:eta} is a generalisation of what was proposed in \cite{fil_separating_2020}, and now includes recombination:

\begin{equation}
    \eta_{RI, rec} = \frac{\int_{\Delta}(S_{ion,lotgt}+S_{ion,recomb})}{\Gamma_{target,tot}-S_{rec}} 
    \label{eq:eta}
\end{equation}

where $\Delta$ is the domain (selected flux tubes) of interest. $S_{ion, lotgt}$ and $S_{ion,recomb}$ are the ionization sources integrated over $\Delta$, due to neutrals generated at the lower outer target and neutrals generated through recombination processes, respectively. The domain $\Delta$ can be the whole lower outer divertor, a single flux tube, or a bundle of flux tubes. In our case, $\Delta$ corresponds to all the flux tubes of the outer divertor (from X-point to target) which are plotted in Figure \ref{fig:grids}, PFR included. Those flux surfaces carry all the heat flux to the divertor target (as can be seen on Figure \ref{fig:tgt} c), f) and i)). In our case, $\eta_{RI, rec}$ can be interpreted as the probability of all neutrals recycling from the lower outer target and being generated by recombination, to be ionized in the lower outer divertor region; $1-\eta_{RI, rec}$ corresponds to the fraction of recycled + recombined neutrals that escape the lower outer divertor - either to the inner divertor or out of the divertor completely (including to the pumps).\\
Figure \ref{fig:flux}b) shows that the neutral trapping is similar between Conventional and Super-X configurations, which is consistent with the fact that $N_{thres,SOLPS} \simeq N_{thres,2PM}$. In other words, magnetic geometry effects (total flux expansion) are dominant over neutral effects in differentiating the two MAST-U configurations, due to the closure of the chamber. Note that this is consistent with the "divertor closure synthetic measurements" reported in Figure 10 of \cite{havlickova_solps_2015-1}, which showed only a small improvement of the closure between the MAST-U conventional and the Super-X (and a very large effect on the closure when removing the MAST-U baffle). Additionally, the MAST-U neutral trapping calculation used in this paper (see equation \ref{eq:eta}) is found to be higher than what was obtained for TCV, i.e. $\eta_{RI,rec} \simeq 0.9$ for MAST-U at rollover while it was $<0.8$ in the TCV unbaffled cases modelled in \cite{fil_separating_2020} (and reprocessed with $\Delta =$ SOL+PFR for consistency). Compared to TCV, MAST-U is indeed much more tightly baffled at the divertor entrance, which facilitates the trapping of neutrals in the divertor. Reprocessing of the TCV modeling with idealized baffles done in \cite{fil_separating_2020} shows an increase of the neutral trapping in those cases, to values close to MAST-U's, which is consistent with the result of recent TCV experiments with baffles \cite{fevrier_divertor_2021}\cite{reimerdes_initial_2021}.

\subsection{Influence of the strike point angle}
\label{sec:beta}
\begin{figure}[!ht]
\centering
\includegraphics[width=15cm]{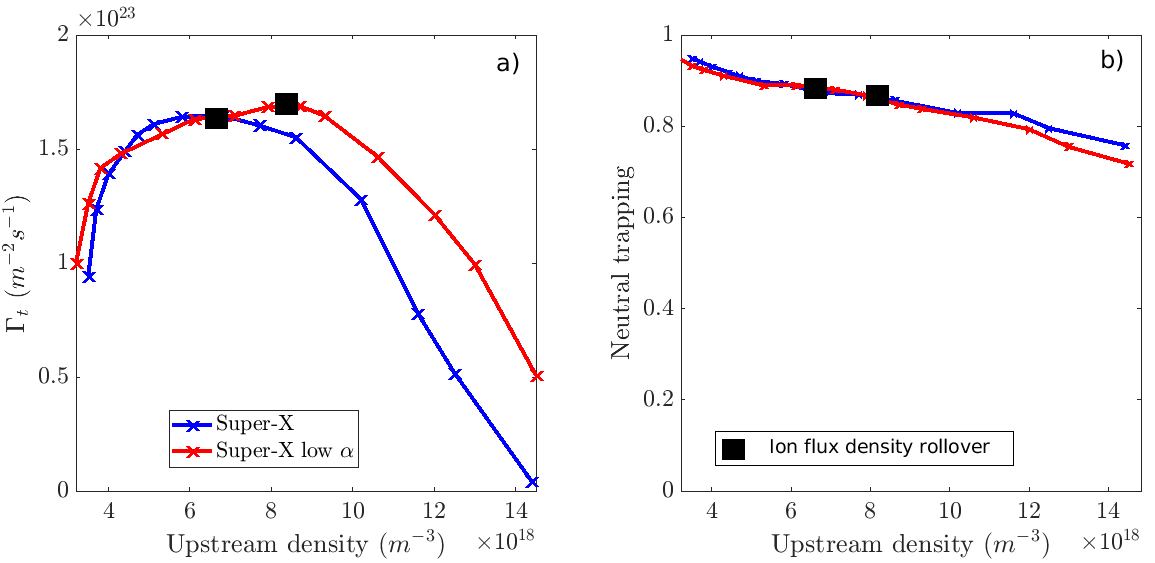}
\caption{Modelled MAST-U Super-X outer lower target ion flux density (a) and associated neutral trapping (defined in equation \ref{eq:eta}) (b), for two values of the strike point angle, $\alpha$ ($162^{\circ}$ vs. $105^{\circ}$).}
\label{fig:beta} 
\end{figure}

In our previous modelling of neutral trapping and total flux expansion for TCV, we demonstrated that the separatrix flux surface angle to the target surface, $\alpha$, can have a significant effect on neutral trapping and on the detachment threshold. In order to study this effect in MAST-U, a second Super-X equilibrium has been generated for which $\alpha$ has been reduced compared to the Super-X studied in the previous section, from $162^{\circ}$ to $105^{\circ}$; i.e. the separatrix is closer to the target normal for this new Super-X "low $\alpha$" equilibrium. The new case, which can be seen in Figure \ref{fig:grids}, is less of a "horizontal target" \cite{lipschultz_neutrals_2007}, meaning that a larger fraction of the recycling neutral trajectories are aimed back towards the separatrix instead of launched further into the SOL (typical of a 'horizontal' target). One would expect for the flux surfaces more normal to the surface (`low $\alpha$') that the detachment threshold would be reduced through more ionization and thus power losses. That same ionization would maximize neutral trapping in that case. \\
In MAST-U, the variation of the separatrix poloidal angle to the surface is found to engender less effect on $\eta_{RI,rec}$ and detachment threshold than what has been observed in TCV.\\
As can be seen on Figure \ref{fig:beta}b), the Super-X with a lower $\alpha$ has essentially the same neutral trapping as the higher $\alpha$ version. It appears that the MAST-U divertor chamber reduces the effect of varying $\alpha$ because its geometry is much more closed than TCV and other existing tokamaks; we conjecture that the length of the chamber as well as the small aperture to the core plasma  traps the neutrals so well that the strike point angle becomes a second order effect (i.e. $\eta_{RI, rec}$ close to 100\%); the neutral trapping is very large, such that very few neutrals escape the leg and chamber for both low- and high-$\alpha$. This is in contrast to TCV \cite{fil_separating_2020}. Despite having the same neutral trapping, the low-$\alpha$ case has a slightly higher density detachment threshold, as can be seen on Figure \ref{fig:beta}a). This is perhaps due to the low-$\alpha$ case having a slightly lower total flux expansion (i.e. smaller outer strike point radius) compared to the higher $\alpha$, as can be seen on Figure \ref{fig:grids}. It is more likely to be due to the low-$\alpha$ case having a significantly smaller connection length than the high-$\alpha$ case (which could lead to a higher detachment threshold).

\section{Further discussion and conclusions}
\label{sec:conclusions}

In the current study, the Super-X configuration has higher total flux expansion compared to the Conventional configuration, which results in a lower detachment threshold; this result is similar to the result of our TCV modelling, where neutral trapping is kept $\approx$ constant across divertor configurations. The neutral trapping for MAST-U is indeed found to be similar between the Conventional and Super-X configurations, and leads to a $N_{thresh}$ close to that predicted by total flux expansion only. \\
When including other control variables as in \cite{lipschultz_sensitivity_2016}\cite{fil_separating_2020}, i.e. also including the divertor impurity concentration $C_{Z, div}$ and the upstream parallel heat flux $q_{u,\parallel}$, the Super-X configuration still has a lower combined detachment threshold (i.e. $\frac{n_{e,up}\sqrt{C_{Z,div}}}{q_{u,\parallel}^{5/7}}$) than the Conventional, but to a lesser extend (i.e. $C_{thres} \sim 1.3$ instead of $N_{thres} \sim 1.6$). This is mainly due to the Super-X having an higher divertor Carbon concentration at rollover than the Conventional. One difference which could explain the disagreement with the model is that impurity radiation is not the dominant power loss mechanism in our simulations (hydrogenic radiation is) while the model assumes so. A more in-depth comparison between SOLPS-ITER and this simplified model \cite{lipschultz_sensitivity_2016} in full geometry is left for future work, noting that such a study in simpler geometry have already been done \cite{cowley_2022}.
\\
Lowering the strike point angle, $\alpha$, seems to only marginally affect the detachment threshold in MAST-U. That result is in contrast to TCV, possibly because of MAST-U's tightly baffled divertor. While the effect of total flux expansion is clearly observed, this study highlights the need to properly model the whole wall structure and the (kinetic) neutral reflections to get an accurate calculation of the neutral trapping and to be able to evaluate and compare different magnetic configurations. \\
Note that the choice of the domain of analysis $\Delta$ is important. The numbers for $N_{thresh}$ given in the previous sections are very similar whether $\Delta = SOL + PFR$, $\Delta = SOL$ or when $\Delta$ is large enough to include most of the target heat flux. But when $\Delta$ corresponds to a single flux tube or a bundle of only a few flux tubes, the quantitative values of $N_{thres}$ and $\eta_{RI,rec}$ between configurations can change. However, the qualitative variation and implications described in the paper still remain the same. The quantitative differences in $\eta_{RI,rec}$ in that case are presumably due to differences in the ionization radial profiles between configurations, which should be investigated further in future work. \\
Ultimately, further simulations (e.g. with drifts) of ongoing MAST-U experiments will bring even more understanding of the role of $\alpha$ and divertor closure on the divertor detachment threshold and the expansion of the detached region from the target towards the X-point. A more accurate treatment of molecular charge exchange in SOLPS-ITER would also likely lead to particle losses at higher temperatures due to Molecular-Activated-Recombination (MAR), as pointed out in \cite{verhaegh_role_2021}, and may change the differences in particle balance between configurations. By validating such simulations against experiments, we would then be in a position to make predictions for the detachment access windows of future fusion reactors prototypes, such as STEP.\\
For STEP predictive simulations, one might also need to include photon opacity effects, which may undo some of the benefits of the Super-X compared to the conventional divertor. We indeed expect that the use of the Super-X configuration will form a larger cloud of neutrals than the conventional configuration, which would enhance opacity and, as a result, modify ionisation/recombination rates and lower hydrogenic radiative losses.

\section{Acknowledgements}
This work has been carried out within the framework of the EUROfusion Consortium and has received funding from the Euratom research and training programme 2014-2018 and 2019-2020 under grant agreement No 633053 and from the RCUK Energy Programme [grant number EP/T012250/1]. To obtain further information on the data and models underlying this paper please contact publicationsmanager@ukaea.uk. The views and opinions expressed herein do not necessarily reflect those of the European Commission. This work has also received funding from the EPSRC under the grant EP/N023846/1. The research by B.Lipschultz was funded in part by the RCUK Energy Programme (EPSRC grant numbers EP/I501045 and EP/N023846/1). This work has been in part performed under the auspices of the U.S. DoE by LLNL under Contract DE-AC52-07NA27344.

\printbibliography
\end{document}